\documentclass[reqno]{amsart}
\usepackage{amssymb}
\usepackage{hyperref}
\usepackage[pdftex]{graphicx}

\AtBeginDocument{{\noindent\small
\sc{Bulletin of Mathematical Analysis and Applications}\newline
ISSN: 1821-1291, URL: http://www.bmathaa.org\newline
Volume x Issue xxx, Pages XXX.}

\thanks{\copyright 2008 Universiteti i Prishtin\"es, Prishtin\"e, Kosov\"e.}
\vspace{9mm}}

\begin{document}
\title[ Multiplicity of Solutions for LPDEs]
{Multiplicity of Solutions for Linear Partial Differential Equations Using (Generalized) Energy Operators}

\author[J.-P.~Montillet]
{Jean-Philippe Montillet}  

\address{Jean-Philippe Montillet \newline
ESPlab, STI IMT, Ecole Polytechnique de Lausanne, MC A3 301 (Microcity), Rue de la Maladière 71b, CH-2002 Neuchâtel, Switzerland}
\email{jean-philippe.montillet@epfl.ch}

\thanks{Submitted July 13, 2016.}



\begin{abstract}
Families of energy operators and generalized energy operators have recently been introduced in the definition of the solutions of linear Partial Differential Equations (PDEs) with a particular application to the wave equation \cite{JPMontillet2014}. To do so, the author has introduced the notion of energy spaces included in the Schwartz space $\mathbf{S}^-(\mathbb{R})$. In this model, the key is to look at which ones of these subspaces are reduced to $\{0\}$ with the help of energy operators (and generalized energy operators). It leads to define additional  solutions for a nominated PDE. Beyond that, this work intends to develop the concept of $\it{multiplicity}$ of solutions for a linear PDE through the study of these energy spaces (i.e. emptiness). The main concept is that the PDE is viewed as a generator of solutions rather than the classical way of solving the given equation with a known form of the solutions together with boundary conditions. The theory is applied to the wave equation with the special case of the evanescent waves. The work ends with a discussion on another concept, the $\it{duplication}$ of solutions and some applications in a closed cavity.
\end{abstract}

\maketitle
\numberwithin{equation}{section}
\newtheorem{theorem}{Theorem}[section]
\newtheorem{lemma}[theorem]{Lemma}
\newtheorem{proposition}[theorem]{Proposition}
\newtheorem{corollary}[theorem]{Corollary}
\newtheorem*{remark}{Remark}

\section{Introduction}

The energy operator was  initially called the Teager-Kaiser energy operator \cite{Kaiser90} and the family of Teager-Kaiser energy operators in \cite{Maragos1995}. It was first introduced in signal processing to detect transient signals \cite{Dunn} and filtering modulated signals \cite{Bovik93}. After three decades of research, it has shown multiple applications in various areas (i.e. speech analysis\cite{Dunn}, transient signal detection\cite{Kandia2006}, image processing \cite{Cexus},  optic \cite{Salzenstein}, localization \cite{Schasse}). This energy operator is defined as $\Psi_2^-$ in \cite{JPMontillet2010}, \cite{JPMontillet2013}, \cite{JPMontillet2014} and through this work. \cite{JPMontillet2010} defined the conjugate operator $\Psi_2^+$ in order to rewrite the wave equation with these two operators. Furthermore, \cite{JPMontillet2013} defined the family of energy operator $( \Psi_k^-)_{k\in\mathbb{Z}}$ and $( \Psi_k^+)_{k\in\mathbb{Z}}$ in order to decompose the successive derivatives of a finite energy function $f^n$ ($n$ in $\mathbb{Z}^+ -\{0,1\}$) in the Schwartz space.
\cite{JPMontillet2014} introduced the concept of  generalized energy operators. In the same work, it was shown that there is a possible application of the energy operators and generalized energy operators to define new sets of solutions for linear PDEs.
%
\\ This work is the sequel of \cite{JPMontillet2014}. It intends to define the concept of $\it{Energy}$ $\it{ Spaces}$, which are sets (in the Schwartz space) of solutions for a nominated linear PDE. $\mathbf{Theorem}$ $2$ states the mechanism for the functions $f^n$ and $\partial_t^k f^n$ to be solutions of a nominated linear PDE. Then, the work is extended to the case when the energy operator families (applied to $f$, $\Psi_p^+(f)$, $\Psi_p^-(f)$) and generalized energy operator families can also be solution of the same equation via the $\mathbf{Corollary}$ $1$. An overview of the concept in the particular case of the wave equation  is to consider these additional solutions as waves generated by a given PDE (i.e. the d' Alembert operator for the wave equation  $\square (f)=0$ \cite{Bryan}) with lower energy. This approach differs with the traditional way of  solving a PDE using boundary conditions with a known form of the solutions. When using the definition of energy space ($\mathbf{Definition}$ $3$, \cite{JPMontillet2014}), the formulation of finding each solution (e.g., $f^n$,  $\partial_t^k f^n$, $\big (\Psi_1^+(f) \big)^n$, $\partial_t^k\big (\Psi_1^+(f) \big)^n$) can be reformulated in looking into which energy space is not reduced to $\{0\}$. Section 6 of \cite{JPMontillet2014} is the proof of concept using the wave equation and the evanescent waves as special type of solutions. The last section concludes with a discussion on the potential of using this model with non-linear PDEs.

\section{Preliminaries}

\subsection{ Energy Operators and Generalized Energy Operators in $\mathbf{S}^-(\mathbb{R})$ }\label{preliminariesSection}

Throughout this work, $f^n$ for any $n$ in $\mathbb{Z}^+-\{0\}$ is supposed to be a smooth real-valued and finite energy function, and in the Schwartz space $\mathbf{S}^{-}(\mathbb{R})$ defined as:
\begin{equation}\label{SRRRRR}
\mathbf{S}^{-}(\mathbb{R}) =\{f \in \mathbf{C}^{\infty}(\mathbb{R}), \qquad {sup}_{t<0} |t^k||\partial_t^j f(t)|<\infty,\qquad \forall k \in \mathbb{Z}^+, \qquad \forall j \in \mathbb{Z}^+ \}
\end{equation}
Sometime $f^n$ can also be analytic if its development in Taylor-Series is necessary (e.g, the application to closed cavity in Section \ref{waveguideappli}). The choice of $f^n$ (for any $n$ in $\mathbb{Z}^+-\{0,1\}$) in the Schwartz space $\mathbf{S}^{-}(\mathbb{R})$ is based on the development in \textit{\cite{JPMontillet2013}, Section 2},  because we are dealing with multiple integrals or derivatives of $f^n$ when applying the energy operators $({\Psi}_{k}^{-})_{k\in\mathbb{Z}^+}$, $({\Psi}_{k}^{+})_{k\in\mathbb{Z}^+}$ and later on the generalized energy operators. 
\\ In the following, let us call the set $\mathcal{F}(\mathbf{S}^{-}(\mathbb{R}),\mathbf{S}^{-}(\mathbb{R}))$ all functions/operators defined such as $\gamma:$ $\mathbf{S}^{-}(\mathbb{R})$ $\rightarrow$ $\mathbf{S}^{-}(\mathbb{R})$. Let us recall some definitions and important results given in \cite{JPMontillet2013} and \cite{JPMontillet2014}.
\vspace{1.0em}
\newline Section $2$ in \textit{\cite{JPMontillet2013}} and Section $4$ in \textit{\cite{JPMontillet2014}} defined the energy operators $\Psi_k^+$, $\Psi_k^-$ ($k$ in $\mathbb{Z}$) and the generalized energy operators $[[.]^p]_k^+$ and $[[.]^p]_k^-$ ($p$ in $\mathbb{Z}^+$). \cite{JPMontillet2014} defined:
\begin{eqnarray}\label{bracketl}
\Psi_k^+(.) &=& \partial_t.\partial_t^{k-1}. + . \partial_t^k., \hspace{0.5em} k\in\mathbb{Z} \nonumber\\
{[.,.]}_k^+ &=& \Psi_k^+(.) \nonumber\\
{[.,.]}_k^+ &=& [.]_k^+ \nonumber 
\end{eqnarray}
$\Psi_k^-$ is the operator conjugate of $\Psi_k^+$.
Furthermore, \cite{JPMontillet2014} defined the generalized energy operators $[[.]^1]_{k}^+$ and $[[.]^1]_{k}^-$ :
\begin{eqnarray}\label{generalizedEO}
{[[.,.]_k^+,[.,.]_k^+]}_{k}^+ &=& \partial_t {\Psi}_k^+(.) \partial_t^{k-1} {\Psi}_k^+(.) + {\Psi}_k^+(f) \partial_t^k {\Psi}_k^+(.) \nonumber \\
{[[.,.]_k^+,[.,.]_k^+]}_{k}^+ &=& \partial_t [[.]^0]_{k}^+ \partial_t^{k-1} [[.]^0]_{k}^+ + [[.]^0]_{k}^+ \partial_t^k [[.]^0]_{k}^+ \nonumber \\
&=&[[.]^1]_{k}^+ \nonumber \\
{[[.,.]_k^-,[.,.]_k^-]}_{k}^- &=& \partial_t {\Psi}_k^-(.) \partial_t^{k-1} {\Psi}_k^-(.) - {\Psi}_k^-(.) \partial_t^k {\Psi}_k^-(.) \nonumber \\
{[[.,.]_k^-,[.,.]_k^-]}_{k}^- &=& \partial_t [[.]^0]_{k}^- \partial_t^{k-1} [[.]^0]_{k}^- + [[.]^0]_{k}^- \partial_t^k [[.]^0]_{k}^- \nonumber \\
&=&[[.]^1]_{k}^- \nonumber \\
\end{eqnarray} 
By $\it{iterating}$ the bracket $[.]$, \cite{JPMontillet2014} defined the generalized operator  $[[.]^p]_{k}^-$ and the conjugate $[[.]^p]_{k}^+$ with $p$ in $\mathbb{Z}^+$. Note that $[[f]^p]_{1}^- =0$ for all $p$ in $\mathbb{Z}^+$.
\newline In addition, the derivative chain rule property and bilinearity of the energy operators and generalized operators are shown respectively in \textit{\cite{JPMontillet2013}, Section 2} and \textit{\cite{JPMontillet2014}, Proposition 3}. 
\vspace{1.0em}
\\$\bold{Definition}$ $1$ \textit{\cite{JPMontillet2013}}: For all $f$ in $\mathbf{S}^{-}(\mathbb{R})$, for all $v\in\mathbb{Z}^+-\{0\}$, for all  $n\in\mathbb{Z}^+$ and $n>1$, the family of operators $(\Psi_k)_{k \in \mathbb{Z}}$ (with $(\Psi_k)_{k \in \mathbb{Z}}$ $\subseteq$ $\mathcal{F}(\mathbf{S}^{-}(\mathbb{R}),\mathbf{S}^{-}(\mathbb{R}))$) decomposes $\partial_t^v$$f^n$ in $\mathbb{R}$, if it exists $(N_j)_{j\in \mathbb{Z}^+ \cup \{0\}}$ $\subseteq$ $\mathbb{Z^+}$,  $(C_i)_{i=-N_j}^{N_j}$ $\subseteq$ $\mathbb{R}$, and it exists $(\alpha_j)$ and $l$ in $\mathbb{Z^+}\cup\{0\}$ (with $l<v$) 
\\ such as $\partial_t^v$$f^n = \sum_{j=0}^{v-1} \big(_{j}^{v-1} \big) \partial_t^{v-1-j} f^{n-l} \sum_{k=-N_j}^{N_j} C_k \Psi_k(\partial_t^{\alpha_k}f)$.
\vspace{1.0em}
\\ In addition, one has to define $\mathbf{s}^{-}(\mathbb{R})$ as:
\begin{equation}
\mathbf{s}^{-}(\mathbb{R}) = \{ f \in  \mathbf{S}^{-}(\mathbb{R}) | f \notin  (\cup_{k \in \mathbb{Z}} Ker(\Psi^{+}_k))\cup(\cup_{k \in \mathbb{Z}-\{1\}} Ker(\Psi^{-}_k))\}
\end{equation}
$Ker(\Psi^{+}_k)$ and $Ker(\Psi^{-}_k)$ are the kernels of the operators $\Psi^{+}_k$ and $\Psi^{-}_k$ ($k$ in $\mathbb{Z}$) (see \textit{\cite{JPMontillet2013}, Properties 1 and 2}). One can also underline that $\mathbf{s}^{-}(\mathbb{R}) \subsetneq \mathbf{S}^{-}(\mathbb{R})$    Following $\bold{Definition}$ $1$, the \emph{uniqueness} of the decomposition  in $\mathbf{s}^{-}(\mathbb{R})$ with the families of differential operators can be stated as:
\vspace{1.0em}
\newline $\bold{Definition}$ $2$ \textit{\cite{JPMontillet2013}}: For all $f$ in $\mathbf{s}^{-}(\mathbb{R})$, for all $v\in\mathbb{Z}^+-\{0\}$, for all  $n\in\mathbb{Z}^+$ and $n>1$, the families of operators $(\Psi^{+}_k)_{k \in \mathbb{Z}}$ and $(\Psi^{-}_k)_{k \in \mathbb{Z}}$ ($(\Psi^{+}_k)_{k \in \mathbb{Z}}$ and $(\Psi^{-}_k)_{k \in \mathbb{Z}}$ $\subseteq$ $\mathcal{F}(\mathbf{s}^{-}(\mathbb{R}),\mathbf{S}^{-}(\mathbb{R}))$) decompose uniquely $\partial_t^v$ $f^n$ in $\mathbb{R}$, if for any family of operators $(S_k)_{k \in \mathbb{Z}}$ $\subseteq$ $\mathcal{F}(\mathbf{S}^{-}(\mathbb{R}),\mathbf{S}^{-}(\mathbb{R})$) decomposing  $\partial_t^v$ $f^n$ in $\mathbb{R}$, there exists a unique couple $(\beta_1,\beta_2)$ in $\mathbb{R}^2$ such as: 
\begin{equation}
S_k(f) = \beta_1 \Psi^{+}_k(f) + \beta_2 \Psi^{-}_k(f), \qquad \forall k\in\mathbb{Z} 
\end{equation}
\vspace{1.0em}
\\ Two important results shown in \textit{\cite{JPMontillet2013} (Lemma and Theorem)} are: 
\vspace{1.0em}
\\$\bold{Lemma}$ $0$: For $f$ in $\mathbf{S}^{-}(\mathbb{R})$, the family of DEO ${\Psi}_{k}^{+}$ ($k=\{0,\pm 1,\pm 2,...\}$) decomposes the successive derivatives of the $n$-th power of $f$ for $n\in\mathbb{Z}^+$ and $n>1$. 
\vspace{1.0em}
\\$\bold{Theorem}$ $0$: For $f$ in $\mathbf{s}^{-}(\mathbb{R})$, the families of DEO ${\Psi}_{k}^{+}$ and ${\Psi}_{k}^{-}$ ($k=\{0,\pm 1,\pm 2,...\}$) decompose uniquely the successive derivatives of the $n$-th power of $f$ for $n\in\mathbb{Z}^+$ and $n>1$. 
\vspace{1.0em}
\\ The $\bold{Lemma}$ $0$ and $\bold{Theorem}$ $0$  were then extended in \textit{\cite{JPMontillet2014}} to the family of generalized operator with :
\vspace{1.0em}
\newline $\bold{Lemma}$ $1$: For $f$ in $\mathbf{S}_{p}^{-}(\mathbb{R})$, $p$ in  $\mathbb{Z}^+$, the families of generalized energy operators $[[.]^{p}]_k^+$ ($k=\{0,\pm 1,\pm 2,...\}$) decompose the successive derivatives of the $n$-th power of $[[f]^{p-1}]_1^+$ for $n\in\mathbb{Z}^+$ and $n>1$.
\vspace{1.0em}
\newline $\bold{Theorem}$ $1$: For $f$ in $\mathbf{s}_p^{-}(\mathbb{R})$, for $p$ in $\mathbb{Z}^+$, the families of generalized operators $[[.]^p]_{k}^+$ and $[[.]^p]_{k}^-$ ($k=\{0,\pm 1,\pm 2,...\}$) decompose uniquely the successive derivatives of the $n$-th power of $[[f]^{p-1}]_{1}^+$ for $n\in\mathbb{Z}^+$ and $n>1$. 
\vspace{1.0em}
\\ $\mathbf{S}_{p}^{-}(\mathbb{R})$ and $\mathbf{s}_{p}^{-}(\mathbb{R})$ ($p$ in $\mathbb{Z}^+$) are energy spaces in $\mathbf{S}^-(\mathbb{R})$ defined in the next section.
\\ Note that as underlined in \cite{JPMontillet2013} (Section 3, p.74) and \cite{JPMontillet2014} (Section $4$), one can extend the $\bold{Theorem}$ $0$, $\bold{Theorem}$ $1$, $\bold{Lemma}$ $0$ and $\bold{Lemma}$ $1$ for $f^n$ with $n$ in $\mathbb{Z}$. The appendix $A$ recalls the discussion in  \cite{JPMontillet2014} (Section $4$). Here, $n$ is restricted to $\mathbb{Z}^+ -\{0,1\}$ in order to easy the whole mathematical development. 
\\  Some time, the finite energy function $f^n$ ($n$ in $\mathbb{Z}^+ -\{0,1\}$) can also be considered analytic. In other words, there are ($p$,$q$) ($p >q$) in $\mathbb{R}^2$ such as $f^n$ can be developed in Taylor Series \cite{Kreizig2003}:
\begin{eqnarray}\label{f2eq}
f^n(p) &=& f^n(q) + \sum_{k=1}^\infty \partial_t^k f^n(q) \frac{(p-q)^k}{k!} \nonumber \\
\end{eqnarray}
\textit{Proposition 1, \cite{JPMontillet2014}} states: 
\vspace{0.5em}
\newline \textit{$\bold{Proposition}$ $1$: If for any $n$ $\in$ $\mathbb{Z}^+$, $f^n$ $\in$ $\mathbf{S}^{-}(\mathbb{R})$ is analytic and finite energy; for any ($p$,$q$) $\in$ $\mathbb{R}^2$ (with $\tau$ in $[q,p]$) and $\mathcal{E}(f^n)$ in $\mathbf{S}^{-}(\mathbb{R})$ is analytic, where}
\begin{equation}\label{EnergyfunDefine02}
\mathcal{E}(f^n(\tau)) = \int_q^{\tau} (f^n(t))^2dt < \infty
\end{equation}
\textit{then} 
\begin{eqnarray}\label{refeq1}
\mathcal{E}(f^n(p)) &=& \mathcal{E}(f^n(q)) + \sum_{k=0}^\infty \partial_t^k (f^n(q))^2 \frac{(p-q)^k}{k!} <\infty \nonumber \\
\end{eqnarray}
\textit{is a convergent series.}
\vspace{0.5em}
\\This property is specifically used in the applications of this work described in Section \ref{Section00Application}.
\subsection{Energy Spaces}
The energy spaces were introduced in \textit{\cite{JPMontillet2014}, Definition 3} and \textit{\cite{JPMontillet2014}, equation (24)}. The definition reads:
\vspace{1.0em}
\newline $\bold{Definition}$ $3$: The energy space $\mathbf{E}_p$, with $p$ in $\mathbb{Z}^+$, is equal to $\mathbf{E}_p=\bigcup_{i\in\mathbb{Z}^+ \cup \{0\}} \mathbf{M}^i$. 
\vspace{1.0em}
\newline with $\mathbf{M}^i$ $\subsetneq \mathbf{S}^{-}(\mathbb{R})$  for $i$ in $\mathbb{Z}^+$ defined 
\begin{equation}\label{energyspaceH}
\mathbf{M}^i=\{ g \in \mathbf{S}^{-}(\mathbb{R}) | \hspace{0.5em} g = \partial_t^i \big ( \big[ [f]^p \big ]_1^+ \big )^n , \big[ [f]^p \big ]_1^+ \in \mathbf{S}^{-}(\mathbb{R}), \hspace{0.5em} n\in\mathbb{Z}^+-\{0,1\}\}
\end{equation}
If $g$ is a general solution of some linear PDEs, then $f^n$ can be identified as a special form of the solution (conditionally to its existence). One can further define the subspace  $\mathbf{S}_p^{-}(\mathbb{R}) \subsetneq \mathbf{S}^{-}(\mathbb{R})$ for $p$ in $\mathbb{Z}^+$ (see \textit{\cite{JPMontillet2014}, equation (25)}):
\begin{equation}
\mathbf{S}_p^{-}(\mathbb{R})=\{ \mathbf{E}_p = \bigcup_{i\in\mathbb{Z}^+ \cup \{0\}} \mathbf{M}^i \neq \{0\} \}
\end{equation}
The energy space $\mathbf{E}_p$ is said  $\emph{associated}$ with $\mathcal{E}( \big[ [.]^{p+1} \big ]_1^+)$. Note that $\mathbf{E}_p$  is not empty, because the assumption is that $\big ( \big[ [f]^p \big ]_1^+ \big )^n$ is a solution of a given linear PDE throughout this work. Thus, the energy space cannot be defined without a nominated PDE.
Now, one can define the subset $\mathbf{s}_p^{-}(\mathbb{R})$ defined as:
\begin{equation}
\mathbf{s}_p^{-}(\mathbb{R}) = \{ f \in  \mathbf{S}_p^{-}(\mathbb{R}) | f \notin  (\cup_{k \in \mathbb{Z}} Ker([[f]^{p}]_k^+)\cup(\cup_{k \in \mathbb{Z}-\{1\}} Ker([[f]^{p}]_k^-))\}
\end{equation}
The subset $\mathbf{s}_p^{-}(\mathbb{R})$ is also defined such as $\mathbf{E}_p\neq \{0\}$. Thus, one can see that $\mathbf{s}_p^{-}(\mathbb{R})\subsetneq\mathbf{S}_p^{-}(\mathbb{R})$. Note that  $\mathbf{s}_0^{-}(\mathbb{R})$ $\subsetneq$ $\mathbf{s}^{-}(\mathbb{R})$, because $\mathbf{S}_p^{-}(\mathbb{R})$ $\subsetneq$ $\mathbf{S}^{-}(\mathbb{R})$ ($p$ in $\mathbb{Z}^+$).
Furthermore, one can also define the special case: 
\begin{equation}\label{energyspaceH00}
\mathbf{N}^i=\{ g \in \mathbf{S}^{-}(\mathbb{R}) | \hspace{0.5em} g = \partial_t^i \big ( f )^n , f^n \in \mathbf{S}^{-}(\mathbb{R}), \hspace{0.5em} n\in\mathbb{Z}^+-\{0,1\}\}
\end{equation}
Following Section $4$ in \cite{JPMontillet2014}, $\mathbf{E} = \bigcup_{i\in\mathbb{Z}^+ \cup \{0\}} \mathbf{N}^i \neq \{0\}$, $\mathbf{E}\subsetneq \mathbf{S}^{-}(\mathbb{R})$. 
%
Finally, the energy space $\mathbf{E}$ is said  $\emph{associated}$ with $\mathcal{E}( \big[ [.]^0 \big ]_1^+)$.  
\section{Theorems of  Multiplicity of the Solutions for a given PDE}
\subsection{Multiplicty of the Solutions }
To recall the first section, a possible application of the theory of the energy operator is to look at the solutions for various values of $n$ and $i$ instead of solving the equation for specific values (e.g., boundary conditions). According to Equation \eqref{energyspaceH} and $\bold{Definition}$ $3$, it is equivalent to find which subspace $\mathbf{M}^i$ ($i\in\mathbb{Z}^+$) of $\mathbf{E}_p$ ($p$ in $\mathbb{Z}^+$) is reduced to $\{0\}$ (or respectively $\mathbf{N}^i$ of $\mathbf{E}$).  Thus using $\bold{Theorem}$ $0$ and $\bold{Theorem}$ $1$, one can use the energy operator in order to find the subspaces $\mathbf{N}^i$ or $\mathbf{M}^i$ reduced to $\{0\}$. For example, let us define for $i$ in $\mathbb{Z}^+$
\begin{equation}\label{energyspaceHb}
\mathbf{L}^{i}=\{ g \in \mathbf{S}^{-}(\mathbb{R})| \hspace{0.5em} g = \partial_t^{i} f^2 = \partial_t^{{i_1}-1} ({\Psi}_{1}^{+} (f) +{\Psi}_{1}^{-} (f)),  \vspace{0.5em} f \in \mathbf{S}^{-}(\mathbb{R}) \}
\end{equation}
with Equation \eqref{energyspaceH}, $\mathbf{L}^{i} \subseteq \mathbf{N}^i$. 
If  $|{\Psi}_{1}^{+}(f)| = 0$ then $\mathbf{L}^{i}  = \{0\}$. Using $\bold{Definition}$ $1$ and $\bold{Theorem}$ $0$, one can  write if it exists $i_1$ in $\mathbb{Z}^+$ such as $|\partial_t^{i_1} {\Psi}_{1}^{+}(f)| = 0$, then $\mathbf{L}^{i_1} = \{0\}$. Subsequently for all $i_2$ $ \geq$ $i_1$, $\mathbf{L}^{i_2} = \{0\}$. 
%
%
%
\subsection{Statement of the Theorem on Multiplicity of the Solutions for a linear PDE}
If $f$ is a solution of a linear PDE, let us call this statement $\Delta^2 f=0$. One can summarize the philosophy of multiple solutions generated by a linear PDE :
\vspace{1em}
\newline $\bold{Theorem}$ $2$ :
$f$ in $\mathbf{s}^-(\mathbb{R})$. Then, $\partial_t^i f^n$ ($i$ in  $\mathbb{Z}^+$, $n$ in $\mathbb{Z}^+-\{0,1\}$) is solution for ($t$, $\tau$) in $[a,b]^2$ ($a<b$, $(a,b)$ in $\mathbb{R}^2$) if it is assumed
\begin{enumerate}
\item [1.](general condition to be a solution)  $\Delta^2 \partial_t^i f^n(\tau) =0$
\item [2.] (Solutions in $\mathbf{S}^-(\mathbb{R})$ ), $\partial_t^i f^n$ is a finite energy function such as $\mathcal{E} (\partial_t^i f^n) (\tau) < \infty$
\item [3.] (3 $\Leftrightarrow$ 2) it exists $m_i$ in $\mathbb{R}$, for $i$ in $\mathbb{Z}^+$ such as $m_i $ $=$ $sup_{ \{\forall n \in \mathbb{Z}^+-\{0,1\},\vspace{0.5em} \tau \in [a,b]  \}}$ $(\mathcal{E} (\partial_t^i f^n) (\tau) )$ 
\item [4.] (Superposition of solutions and energy conservation ) $F(\tau) = \sum_{ k\in \mathbb{Z}^+} \partial_t^k f^n(\tau)$, then $\mathcal{E}(F(\tau)) < \infty$
\item [5.] (5 $\Rightarrow$ 4) $\exists$ $i_1$ in  $\mathbb{Z}^+$ such as $\forall$ $i \geq $ $i_1$, $\mathbf{N}^i = \{0\}$   
\end{enumerate}
\vspace{1em}
From the statement (5.), one can then define the energy space $\mathbf{E} = \bigcup_{i\in [0,i_1-1]} \mathbf{N}^i$. Now, if we want to use the decomposition of  the solutions $\partial_t^i f^n$ with the energy operators (e.g,  $\bold{Theorem}$ $0$) in order to find $i_1$ such as $\mathbf{N}^{i_1}$ is reduced to $\{0\}$, then the definition of the subspace $\mathbf{N}^{i}$ reads 
\begin{eqnarray}\label{energyspaceH0022}
\mathbf{N}^i&=&\{ g \in \mathbf{S}^{-}(\mathbb{R}) | \hspace{0.5em} g = \partial_t^i \big ( f )^n = \alpha_n (\partial_t^{i-1} f^{n-2} (\Psi_1^+(f) + \Psi_1^-(f)), \nonumber \\ 
& & f \in \mathbf{s}^{-}(\mathbb{R}), \hspace{0.5em} n\in\mathbb{Z}^+-\{0,1\}, \hspace{0.5em} \alpha_n \in \mathbb{R}\}
\end{eqnarray}
That is why we define $\bold{Theorem}$ $2$ with $f$ in $\mathbf{s}^-(\mathbb{R})$.
Furthermore, $\bold{Theorem}$ $2$ can be extended to the generalized operators as solutions of the linear PDE. With $p$ in $\mathbb{Z}^+$  and the energy space definition (e.g, $\bold{Definition}$ $3$), the $\bold{Corollary}$ $1$ of $\bold{Theorem}$ $2$ reads
\vspace{1em}
\newline $\bold{Corollary}$ $1$:
For $p$ in $\mathbb{Z}^+$, $f$ in $\mathbf{s}_p^-(\mathbb{R})$, $\partial_t^i ([[f]^p]^+_1)^n$ ($i$ in  $\mathbb{Z}^+$, $n$ in $\mathbb{Z}^+-\{0,1\}$) is solution for ($t$, $\tau$) in $[a,b]^2$ ($a<b$, $(a,b)$ in $\mathbb{R}^2$) if
\begin{enumerate}
\item [1.] (general condition to be a solution) for $i$ in $\mathbb{Z}^+$ and  $n$ in $\mathbb{Z}^+-\{0,1\}$, $\Delta^2 \partial_t^i ([[f]^p]^+_1)^n =0$ 
\item [2.] (Solutions in $\mathbf{S}^-(\mathbb{R})$ ) for $i$ in $\mathbb{Z}^+$ and  $n$ in $\mathbb{Z}^+-\{0,1\}$, $\partial_t^i ([[f]^p]^+_1)^n$ is a finite energy function such as $\mathcal{E} (\partial_t^i ([[f]^p]^+_1)^n) (\tau) < \infty$
\item [3.] (3 $\Leftrightarrow$ 2) it exists $m_i$ in $\mathbb{R}$, for $i$ in $\mathbb{Z}^+$ such as $m_i = sup_{ \{\forall n \in \mathbb{Z}^+-\{0,1\}  \}}$ $(\mathcal{E} (\partial_t^i ([[f]^p]^+_1)^n) (\tau) )$ 
\item [4.] (superposition of solutions and energy conservation )
\\$F(\tau) = \sum_{k \in \mathbb{Z}^+} \partial_t^k ([[f]^p]^+_1)^n)(\tau) < \infty$ 
\item [5.] (5 $\Rightarrow$ 4) $\exists$ $i_1$ in  $\mathbb{Z}^+$ such as $\forall$ $i \geq $ $i_1$, $\mathbf{M}^i = \{0\}$   
\end{enumerate}
\vspace{1em}
%
To recall the remark at the end of the statement of $\mathbf{Theorem}$ $2$, one can justify that $f$ in $\mathbf{s}_p^-(\mathbb{R})$ in order to use the decomposition with the generalized energy operators (e.g, $\mathbf{Theorem}$ $1$). Thus, the definition of the subspace $\mathbf{M}^i$ reads
\begin{eqnarray}\label{energyspaceH22vxd}
\mathbf{M}^i&=&\{ g \in \mathbf{S}^{-}(\mathbb{R}) | \hspace{0.5em} g = \partial_t^i \big ( \big[ [f]^p \big ]_1^+ \big )^n = \nonumber \\
&& \alpha_n (\partial_t^{i-1} \big ( \big[ [f]^p \big ]_1^+ \big )^{n-2} (\big[ [f]^{p+1} \big ]_1^+  + \big[ [f]^{p+1} \big ]_1^-), \nonumber \\
&& f \in \mathbf{s}_p^{-}(\mathbb{R}), \hspace{0.5em} n\in\mathbb{Z}^+-\{0,1\}, \hspace{0.5em} \alpha_n \in \mathbb{R}\} \nonumber \\
\end{eqnarray}
\\$\mathbf{Theorem}$ $2$ and $\mathbf{Corollary}$ $1$   could then be applied to a given linear PDE by replacing the first statement ($\Delta^2 (.) =0$). 
The next section highlights the assumptions behind each assertion in $\mathbf{Theorem}$ $2$ and $\mathbf{Corollary}$ $1$.
\subsection{Underlining Hypothesis of the Theorems and Proofs}
This section discusses each statement in $\bold{Theorem}$ $2$ and $\bold{Corollary}$ $1$. We are not  attempting to give a formal proof, because we conjecture the existence of solutions for a given PDE such as the energy spaces $\mathbf{S}_p^-(\mathbb{R})$ and $\mathbf{s}_p^-(\mathbb{R})$ ($p$ in $\mathbb{Z}^+$) are not reduced to $\{\O\}$. Keeping that in mind, one can give the following explanations to understand the previous section.     
\subsubsection{Theorem 2}
%
\begin{proof}
\begin{itemize}
\item [1.] This is the definition of a solution for a nominated PDE.
\item [2.] All the solutions considered in this work are finite energy functions in order to apply the decomposition using the energy operators stated in $\mathbf{Theorem}$ $0$ and $\mathbf{Theorem}$ $1$. Note that it does not rule out that for some solutions (particular values of $i$ and $n$), one can have $(\mathcal{E} (\partial_t^i f^n) (\tau) ) \sim \infty$. In this case, the solutions cannot be accepted. 
\item [3.] The existence of the upper bound of $\mathcal{E} (\partial_t^i f^n) (\tau)$ ( $\tau$ in $[a,b]$, $\forall n \in \mathbb{Z}^+-\{0,1\}$) comes from the definition. However, let us define the subspaces $H_i$ ($i$ in $\mathbb{Z}^+$),  $H_i$ $\subsetneq$ $\mathbb{R}$ such as
\begin{equation}
H_i = \{ z \in \mathbb{R} | z=\mathcal{E} (\partial_t^i f^n) (\tau) <\infty, \vspace{0.5em} f^n \in \mathbf{S}(\mathbb{R}^-), \vspace{0.5em}  n \in \mathbb{Z}^+-\{0\}, \vspace{0.5em} \tau \in [a,b]\} 
\end{equation}
One can then define $m_i$ in $\mathbb{R}$ \cite{UCDavis} such as
\begin{equation}
m_i = sup_{ \{\forall n \in \mathbb{Z}^+-\{0,1\}, \tau \in [a,b]  \}}(\mathcal{E} (\partial_t^i f^n) (\tau) )
\end{equation}
Furthermore, we can also define $M$ such as 
\begin{equation}
M = max_{\forall i \in \mathbb{Z}^+-\{0\}}(m_i)
\end{equation}
Then, 
\begin{equation}
M = sup_{ \{\forall n \in \mathbb{Z}^+-\{0,1\}, \tau \in [p,q],  \forall i \in \mathbb{Z}^+ \}}(\mathcal{E} (\partial_t^i f^n) (\tau) )
\end{equation}
\item [4.] This statement is to guarantee that there is a finite sum of energy. In other words, there is no infinite number of solutions. Thus with the development in statement (3.), one can use the Minkowski inequality (e.g,  \cite{Hardy}, Theorem 202) for $\tau$ in $[a,b]$
\begin{eqnarray}
\mathcal{E} (F(\tau)) &=& \int_a^\tau |\sum_{ i\in \mathbb{Z}^+} \partial_t^i f^n(t)|^2 dt \nonumber \\
\big ( \mathcal{E} (F(\tau)) \big )^{0.5} & \leq & \sum_{ i\in \mathbb{Z}^+} \big ( \int_a^\tau |\partial_t^i f^n(t)|^2 dt \big )^{0.5} \nonumber \\
\big ( \mathcal{E} (F(\tau)) \big )^{0.5} & \leq & \sum_{ i\in \mathbb{Z}^+} m_i^{0.5}
\end{eqnarray} 
Thus, (4.) stands if and only if $\sum_{ i\in \mathbb{Z}^+} m_i^{0.5} < \infty$. As for all $i\in \mathbb{Z}^+$, $m_i$ is in $\mathbb{R}^+$, it then exists $i_o$ in $\mathbb{Z}^+$ such as $\forall i > i_o$, then $m_i =0$.
\item [5.] Following the discussion in Section $2.2$ (e.g, $\bold{Definition}$ $3$), one can see that if it exists $i_1$ such as  for all $i\geq i_1$, $\mathbf{N}^i$ $= $  $\{0\}$ then   $\mathcal{E} (\partial_t^i f^n)$ $=$ $0$. Or with the above development in (4.), the sum of coefficients $m_i^{0.5}$ becomes finite:
\begin{eqnarray}
\big ( \mathcal{E} (F(\tau)) \big )^{0.5} & \leq & \sum_{ i\in \mathbb{Z}^+} m_i^{0.5} \nonumber \\
\big ( \mathcal{E} (F(\tau)) \big )^{0.5} & \leq & \sum_{ i =0}^{i_1-1} m_i^{0.5}
\end{eqnarray} 
\end{itemize}
\end{proof}
%
%
\subsubsection{Corollary 1}
\begin{proof}
The structure of the $\bold{Corollary}$ $1$ is very similar to $\bold{Theorem}$ $2$. In order to avoid repetitions, some explanations are shortened. For $p$ in $\mathbb{Z}^+$,  
\begin{itemize}
\item [1.] This is the definition of a solution for a nominated PDE.
\item [2.] All the solutions considered in this work are finite energy functions in order to apply the decomposition using the energy operators stated in $\mathbf{Theorem}$ $0$ and $\mathbf{Theorem}$ $1$.
\item [3.] Similarly as in the previous discussion, the existence of the upper bound of $\mathcal{E} (\partial_t^i ([[f]^p]^+_1)^n) (\tau)$ ( $\tau$ in $[a,b]$, $\forall n \in \mathbb{Z}^+-\{0,1\}$) comes from the definition. Similarly to the statement (3.) of $\bold{Theorem}$ $2$, one  can define $H^p_i$ ($i$ in $\mathbb{Z}^+$),  $H^p_i \subsetneq \mathbb{R}$ such as
\begin{equation}
H^p_i = \{ z \in \mathbb{R} | z=\mathcal{E} (\partial_t^i ([[f]^p]^+_1)^n) (\tau) <\infty, \vspace{0.5em} f^n \in \mathbf{S}(\mathbb{R}^-), \vspace{0.5em}  n \in \mathbb{Z}^+-\{0\}, \vspace{0.5em} \tau \in [a,b]\} 
\end{equation}
Subsequently, with $m_{i,p}$ in $\mathbb{R}$ 
\begin{equation}
m_{i,p} = sup_{ \{\forall n \in \mathbb{Z}^+-\{0,1\}, \tau \in [a,b]  \}}(\mathcal{E} (\partial_t^i ([[f]^p]^+_1)^n) (\tau) )
\end{equation}
and defining $M$ such as 
\begin{equation}
M_p = max_{\forall i \in \mathbb{Z}^+-\{0\}}(m_{i,p})
\end{equation}
Then, 
\begin{equation}
M_p = sup_{ \{\forall n \in \mathbb{Z}^+-\{0,1\}, \tau \in [p,q],  \forall i \in \mathbb{Z}^+ \}}(\mathcal{E} (\partial_t^i ([[f]^p]^+_1)^n) (\tau) )
\end{equation}
Following statement [3.] and the previous discussion, one can use the Minkowski inequality (e.g, \cite{Hardy}, Theorem 202) for $\tau$ in $[a,b]$
\begin{eqnarray}
\mathcal{E} (F(\tau)) &=& \int_a^\tau |\sum_{ i\in \mathbb{Z}^+} \partial_t^i ([[f(t)]^p]^+_1)^n|^2 dt \nonumber \\
\big ( \mathcal{E} (F(\tau)) \big )^{0.5} & \leq & \sum_{ i\in \mathbb{Z}^+} \big ( \int_a^\tau |\partial_t^i ([[f(t)]^p]^+_1)^n|^2 dt \big )^{0.5} \nonumber \\
\big ( \mathcal{E} (F(\tau)) \big )^{0.5} & \leq & \sum_{ i\in \mathbb{Z}^+} {m_{i,p}}^{0.5}
\end{eqnarray} 
As previously underlined, (4.) stands if and only if $\sum_{ i\in \mathbb{Z}^+} {m_{i,p}}^{0.5} < \infty$. In other words, it then exists $i_o$ in $\mathbb{Z}^+$ such as $\forall i > i_o$, then $m_{i,p} =0$.
\item [5.] Following previous discussions for $\mathbf{Theorem}$ $2$ and the above development, the sum of coefficients $m_{i,p}^{0.5}$ becomes finite:
\begin{eqnarray}
\big ( \mathcal{E} (F(\tau)) \big )^{0.5} & \leq & \sum_{ i\in \mathbb{Z}^+} m_{i,p}^{0.5} \nonumber \\
\big ( \mathcal{E} (F(\tau)) \big )^{0.5} & \leq & \sum_{ i =0}^{i_1-1} m_{i,p}^{0.5}
\end{eqnarray} 
\end{itemize}
\end{proof}
\subsection{Discussion on non-linear PDEs}\label{nonlinearpdes}
The current work started with the preliminary study in \cite{JPMontillet2010} where the energy operators showed how to decompose the Helmholtz equation. The Helmholtz equation was chosen as an easy application, because of its linearity and the form of the solutions. In \cite{JPMontillet2014}, the preliminary results were extended. However with the additional development in this current work, the introduction of the energy space with the definition based on the solution  of a nominated PDE does not justify anymore the assumption that the PDE must be linear. In addition, the assumption on the linearity of the nominated PDE in $\bold{Theorem}$ $2$ and $\bold{Corollary}$ $1$ does not play any role. Thus, one should also be able to apply this model to non-linear PDEs.  
\section{Application to linear PDEs}\label{Section00Application}
\subsection{Comments on functions of two variables solutions of linear and Nonlinear PDEs}\label{defsectionlPDE}
In this section and the remainder of this work, the finite energy functions of one variable described in Section \ref{preliminariesSection}, are now functions of two variables referring to  the space dimension ($r$) and time ($t$). One has to add in the notation of the operators the symbol $t$ or $r$ to indicate which variable the derivatives refer to. For example,  the operators $\Psi_k^{-,r}(.)$ and $[[.]^p]_k^{+,r}$ ($k$ in $\mathbb{Z}$, $p$ in $\mathbb{Z}^+$) refer to their derivatives in space, whereas $\Psi_k^{-,t}(.)$ and $[[.]^p]_k^{+,t}$ ($k$ in $\mathbb{Z}$, $p$ in $\mathbb{Z}^+$) refer to their derivatives in time. 
Following the  discussion in \cite{JPMontillet2014} (Section 6), one can then define the Schwartz space $\mathbf{S}^{-}(\mathbb{R}^2)$ for function of two variables such as:
\begin{equation}\label{multipleSR}
\begin{split}
\mathbf{S}^{-}(\mathbb{R}^2) =\{f \in \mathbf{C}^{\infty}(\mathbb{R}),  \hspace{0.5em}  \forall (r_0, t_0) \in \mathbb{R}^+| \hspace{0.5em}  {sup}_{t<0} |t^k||\partial_t^j f(r_0, t)|<\infty, \\
& \hspace{-24em} and \hspace{0.5em} {sup}_{r<0} |r^k||\partial_r^j f(r, t_0)|<\infty ,\hspace{0.5em} \forall k \in \mathbb{Z}^+, \hspace{0.5em} \forall j \in \mathbb{Z}^+ \}
\end{split}
\end{equation}
Following this definition, the extension of the subspace $\mathbf{s}^{-}(\mathbb{R}^2) \subseteq \mathbf{S}^{-}(\mathbb{R}^2)$ is:
\begin{eqnarray}
\mathbf{s}^{-}(\mathbb{R}^2) &=& \{f\in \mathbf{S}^{-}(\mathbb{R}^2)|  \hspace{0.2em}\forall\hspace{0.2em} k \in \mathbb{Z}, \hspace{0.2em}  \Psi^{+,t}_k(f) \neq \{0\} \hspace{0.2em}  \nonumber \\
                   & & \hspace{0.2em} \forall\hspace{0.2em} k \in \mathbb{Z}-\{1\}, \hspace{0.2em}\Psi^{-,t}_k(f) \neq \{0\} \} \bigcup \nonumber  \\
                   & & \{f\in \mathbf{S}^{-}(\mathbb{R}^2)|  \hspace{0.2em}\forall\hspace{0.2em} k \in \mathbb{Z}, \hspace{0.2em}  \Psi^{+,r}_k(f) \neq \{0\} \hspace{0.2em}  \nonumber \\
                   & & \hspace{0.2em} \forall\hspace{0.2em} k \in \mathbb{Z}-\{1\}, \hspace{0.2em}\Psi^{-,r}_k(f) \neq \{0\} \}  \nonumber
\end{eqnarray}
In \cite{JPMontillet2014}, it was emphasized that $Definition$ $0$, $Definition$ $1$, $Lemma$ $0$, $Lemma$ $1$, $Theorem$ $0$ and $Theorem$ $1$ can easily be extended to the function of two variables using the above definitions of $\mathbf{s}^{-}(\mathbb{R}^2)$ and $\mathbf{S}^{-}(\mathbb{R}^2)$. We will not state formally all the work previously done for the case of  the functions of two variables in  $\mathbf{S}^{-}(\mathbb{R}^2)$. It is only a matter of replacing the variables from time to space. 
\\ According to the previous section, one can state in the case of a function of two variables that it exists $({\alpha_n}_1, {\alpha_n}_2)$ in $\mathbb{R}^2$ such as for $ f \in \mathbf{s}^{-}(\mathbb{R}^2), \hspace{0.5em} n \in \mathbb{Z}^+ -\{0,1\}$:
\begin{equation}\label{Problemsetup2}
 \left\{
\begin{array}{rl}
\partial_t^i f^n = & {\alpha_n}_1 \Big ( \partial_t^{i-1} \big (f^{n-2} (\Psi_{1}^{t,+}(f)+\Psi_{1}^{t,-}(f))\big) \Big), \\
\partial_r^i f^n = & {\alpha_n}_2 \Big ( \partial_r^{i-1} \big ( f^{n-2} (\Psi_{1}^{r,+}(f)+\Psi_{1}^{r,-}(f)) \big) \Big) \\
\end{array} \right.
\end{equation}
The definition of the energy space $\mathbf{E}$ (e.g, $\bold{Definition}$ $3$) is defined in $\mathbf{S}^{-}(\mathbb{R}^2)$
\begin{eqnarray}\label{energyspaceH002fun}
\mathbf{N}_1^i &=&\{ g \in \mathbf{S}^{-}(\mathbb{R}^2),\hspace{0.5em} \forall t \in \mathbb{R}^+, \hspace{0.5em} r_0 \in \mathbb{R} | g(r_0,t) = \partial_t^i f^n(r_0,t),  \nonumber \\
&& \hspace{0.5em} f  \in \mathbf{s}^{-}(\mathbb{R}^2),  \hspace{0.5em}n \in \mathbb{Z}^+ -\{0,1\} \} \nonumber \\
\mathbf{N}_2^i &=&\{  g \in \mathbf{S}^{-}(\mathbb{R}^2), \forall r \in \mathbb{R}, \hspace{0.5em} t_0 \in \mathbb{R}^+ | g(r,t_0) = \partial_r^i f^n(r,t_0), \nonumber \\
&& \hspace{0.5em} f  \in \mathbf{s}^{-}(\mathbb{R}^2), \hspace{0.5em} n \in \mathbb{Z}^+ -\{0,1\} \} \nonumber 
\end{eqnarray} 
Thus, $\mathbf{E} = \{\bigcup_{i\in\mathbb{Z}^+ \cup \{0\}} \mathbf{N}_1^i \neq \{0\}\} \cup \{\bigcup_{i\in\mathbb{Z}^+ \cup \{0\}} \mathbf{N}_2^i\neq \{0\}\}$, $\mathbf{E}\subsetneq \mathbf{S}^{-}(\mathbb{R}^2)$. Similarly, one can further extend the definition of energy space with the generalized energy operators in $\mathbf{S}_p^{-}(\mathbb{R}^2)$ ($p$ in $\mathbb{Z}^+$) with defining $\mathbf{M}_1^i$ and $\mathbf{M}_2^i$.
\\ Furthermore, it is worth underlining that if $f$ is a solution of a given linear PDE, then $f^n(r_0,t)$ is in  $\mathbf{N}_1^0$ and $f^n(r,t_0)$ is in  $\mathbf{N}_2^0$. Choosing a wise choice for the definition of $f$ (i.e. amplitude not equal to $0$) leads to $\mathbf{N}_1^0$ and $\mathbf{N}_2^0$ not equal to $\{0\}$. One can then conclude that the energy space $\bold{E} \neq \{\O\}$ and $\bold{E} \neq \{0\}$ . 
\subsection{Evanescent waves and the wave equation}
The evanescent waves were already chosen in the numerical example in Section $6$ of \cite{JPMontillet2014}. Here, this type of solutions of the Helmholtz equation is used as an application of the multiplicity theorem stated in $\bold{Theorem}$ $2$. From \cite{Petit} or \cite{Amzallag}, the Helmholtz equation can be formulated :
%
\begin{equation}\label{eqpdep}
 \left\{
\begin{array}{rl}
\partial_r^{2}g(r,t) &  - \frac{1}{c^2} \partial_t^{2}g(r,t) = 0 ,\\
or \hspace{0.5em} \square g(r,t) = 0, \\
t \in [0, T ], \hspace{0.5em} r \in [r_1, r_2], & \hspace{0.5em} (r_1,r_2, T) \in \mathbb{R}^3, \hspace{0.5em} r_1<r_2\\
 t_0 \in [0, T ], \hspace{0.5em} r_0 \in [r_1, r_2] \\
\end{array} \right.
\end{equation}
$c$ is the speed of light. Note that the values of $t$ and $r$ are restricted to some interval, because it is conventional to solve the equation for a restricted time interval in $\mathbb{R}^+$ and a specific region in space. Furthermore, it is well-known that the general solution $g(r,t)$ of this equation is a sum of two waves traveling in opposite direction such as $g(r,t) = f_1(t-r/c)+f_1(t+r/c)$ (e.g., \cite{Petit}). With the development in the previous sections and taking the wave traveling along the increasing positive axis (if the other solutions is considered to travel along the increasing negative axis), we are interested in the solutions of the kind $g(r,t) = \partial_t^i f_1^n(r,t)$ ($n$ in $\mathbb{Z}^+-\{0,1\}$, $p$ in $\mathbb{Z}^+$). 
%
%
%
%
\newline \indent As underlined in  \cite{JPMontillet2014}, solutions in $\mathbf{S}^{-}(\mathbb{R}^2)$ of equation \eqref{eqpdep} are finite energy functions such as the ones decaying for large values of $r$ and $t$. This is a very limiting condition. For example, planar waves are not included.
\\ \indent Now, let us apply the multiplicity theorem (e.g, $\mathbf{Theorem}$ $2$) to the evanescent waves  \cite{Amzallag} :
\begin{equation}\label{evanescent}
 \left\{
\begin{array}{rl}
f(r,t) = & Real \{ A \exp{(k_2 r)} \exp{(j(\omega t -k_1 r))} \} ,\\
t \in [0, T ], \hspace{0.5em} r \in [r_1, r_2], & \hspace{0.5em} (r_1,r_2) \in \mathbb{R}^2, \hspace{0.5em} r_1<r_2\\
\end{array} \right.
\end{equation}
$k_1$ and $k_2$ are the wave numbers,  $\omega$ is the angular frequency and $A$ is the amplitude of this wave \cite{Petit}. Assuming $\omega$ and ($k_1$, $k_2$) known, one can add some boundary conditions in order to estimate $k_1$, $k_2$ and $A$. However, our interest is just the general form assuming that all the parameters are known.
\\ \indent The solutions in $\mathbf{N}^i_1$ and $\mathbf{N}^i_2$ can be stated
\begin{equation}\label{evanescentS1}
 \left\{
\begin{array}{rl}
 \partial_t^i f^n (r_0,t) = & (j\omega n)^i  f^n(r_0,t)\} ,\\
 \partial_r^i f^n (r,t_0) = & (n(k_2-jk_1))^i f^n(r,t_0)\},\\
t \in [0, T ], \hspace{0.5em} r \in [r_1, r_2], & \hspace{0.5em} (r_1,r_2, T) \in \mathbb{R}^3, \hspace{0.5em} r_1<r_2\\
 t_0 \in [0, T ], \hspace{0.5em} r_0 \in [r_1, r_2], \hspace{0.5em} n \in \mathbb{Z}^+ -\{0,1\}, \hspace{0.5em} i \in \mathbb{Z}^+-\{0\} \\
\end{array} \right.
\end{equation}
It is well-known that a traveling wave solution of equation \eqref{eqpdep} should satisfy the dispersion relationship  between $k_1$, $k_2$ and $\omega$  \cite{Petit}. Note that $i$ is in $\in \mathbb{Z}^+-\{0\}$, because we are doing an application where the energy operators should help to determine which energy space is not empty. For $i$ equal $0$, the energy operators are not involved. This space can be seen as the form of desired solutions (with $n$ in $\mathbb{Z}^+ -\{0,1\}$).
\\If we replace the general solution $g$ in equation \eqref{eqpdep} with solutions in $\mathbf{N}^i_1$ and $\mathbf{N}^i_2$ (see Section $4.1$), one can study the dispersion relationship for different solutions of the form $\partial_t^i f^n (r,t)$ or  $\partial_r^i f^n (r,t)$  :
\begin{eqnarray}\label{evanescentS1b}
\square \partial_t^i f^n (r,t) =0 \nonumber \\
Real\{\frac{(j\omega n)^2}{c} - (n(k_2-jk_1))^2\} =0 \nonumber \\
Real\{\frac{(j\omega )^2}{c} - ((k_2-jk_1))^2\} =0  \nonumber \\
\square \partial_r^i f^n (r,t) =0 \nonumber \\
Real\{\frac{(j\omega n)^2}{c} - (n(k_2-jk_1))^2 \}=0 \nonumber \\
Real\{\frac{(j\omega )^2}{c} - ((k_2-jk_1))^2\} =0 
\end{eqnarray}
It means that the dispersion relationship for this type of solutions is a function neither of the degree of the derivatives $i$ nor the power  $n$ for the special case of the evanescent waves. 
Furthermore, one can also calculate the closed-form expression $ \partial_t^i \Psi_1^{+,t} (r,t)$ and $ \partial_r^i \Psi_1^{+,r} (r,t)$ such as
\begin{eqnarray}\label{evanescentS2Psi}
\partial_t^i \Psi_1^{+,t} (r_0,t) &=& Real \{ (j 2 i \omega ) f^2 (r_0,t)\} \nonumber \\
\partial_r^i \Psi_1^{+,r}(r,t_0) &=& Real \{  (2i(k_2-jk_1)) f^2 (r,t_0)\} \nonumber \\
t \in [0, T ], && \hspace{0.5em} r \in [r_1, r_2],  \hspace{0.5em} (r_1,r_2, T) \in \mathbb{R}^3, \hspace{0.5em} r_1<r_2 \nonumber \\
t_0 \in [0, T ], && \hspace{0.5em} r_0 \in [r_1, r_2], \hspace{0.5em} n \in \mathbb{Z}^+ -\{0,1\}, \hspace{0.5em} i \in \mathbb{Z}^+-\{0\}
\end{eqnarray}
These expressions can help to determine which subspaces $\mathbf{N}^i_1$ and $\mathbf{N}^i_2$ can be reduced to $\{0\}$ according to the model explained in Section $3$ (e.g, $\bold{Theorem}$ $1$).
\\ Now, let us estimate numerically $i_1$ such as $|\partial_t^{i_1}\Psi_1^{+,t}(f)| =$ $0$ and $i_2$ such as $|\partial_r^{i_2}\Psi_1^{+,r}(f)| =$ $0$ for fixed values of $A$, $k_1$, $k_2$, $\omega$ (in $\mathbb{R}$)  and a given region in space and interval in time. Note that it is mathematically more accurate to state $|\partial_t^{i_1}\Psi_1^{+,t}(f)| \sim $ $0$ and $|\partial_r^{i_2}\Psi_1^{+,r}(f)| \sim$ $0$ rather than using the symbol '$=$', because we are estimating numerically the different solutions. Thus, one can decide implicitly that it exists $\epsilon$ in $\mathbb{R}$ such as if  $|\partial_t^{i_1}\Psi_1^{+,t}(f)| < \epsilon$ , then $|\partial_t^{i_1}\Psi_1^{+,t}(f)| =0$ (reciprocally $|\partial_r^{i_1}\Psi_1^{+,r}(f)| < \epsilon$ , then $|\partial_r^{i_1}\Psi_1^{+,r}(f)| =0$).
\\ \indent The parameters displayed in Table $1$ are for two scenarios. The parameters in Scenario $1$ are used  in the simulation to estimate $i_1$ and reciprocally Scenario $2$ for $i_2$.
\begin{table}
\caption{Parameters for the scenarios $1$ and $2$}
\begin{tabular}{ | l | l | l |}
   \hline
  Parameters  & Scenario $1$   &  Scenario $2$\\ \hline
    $k_1$ ($m^{-1}$ )& $10^{-10}$ & 0.001  \\ \hline
    $k_2$ ($m^{-1}$ ) & $10^{-10}$ & $10^{-5}$\\ \hline
    $A$ ($m$) & 1    & 1 \\ \hline
    $\omega $ ( $sec^{-1}$) & 0.03 & 3000   \\ \hline
    $T$ ( $sec$) & 100  & 100 \\ \hline
    $[r_1,r_2]$ ( $m$) & $[2, 140]$  & $[2, 140]$ \\ \hline
   \hline
   \end{tabular}
   \end{table}
As previously mentioned in the definition of the energy spaces (e.g, $\bold{Definition}$ $3$), the energy function is said "$\it{associate}$" with the energy space. In other words, one can estimate the energy of $\partial_t^i \Psi_1^{+,t} (f)(r_0,t)$ and $\partial_r^i \Psi_1^{+,r} (f) (r,t_0)$ for the various values of $i$ in $\mathbb{Z}^+-\{0\}$. Figure $1$ and $2$ displays the results. Note that the simulations are performed using Matlab ($v9.0$ $R2016a$ - $mathworks.com$). 
\begin{figure}\label{Fig1}
\caption{ Results from the estimation of the energy of $\partial_t^i \Psi_1^{+,t} (f) (r_0,t)$ and $\partial_t^i [[f]^1]^{+,t}_1 (r_0,t))$.  The Figures correspond to Scenario $1$.}
 \centering
 \hspace{-5em}
   \includegraphics[width=1.3\textwidth]{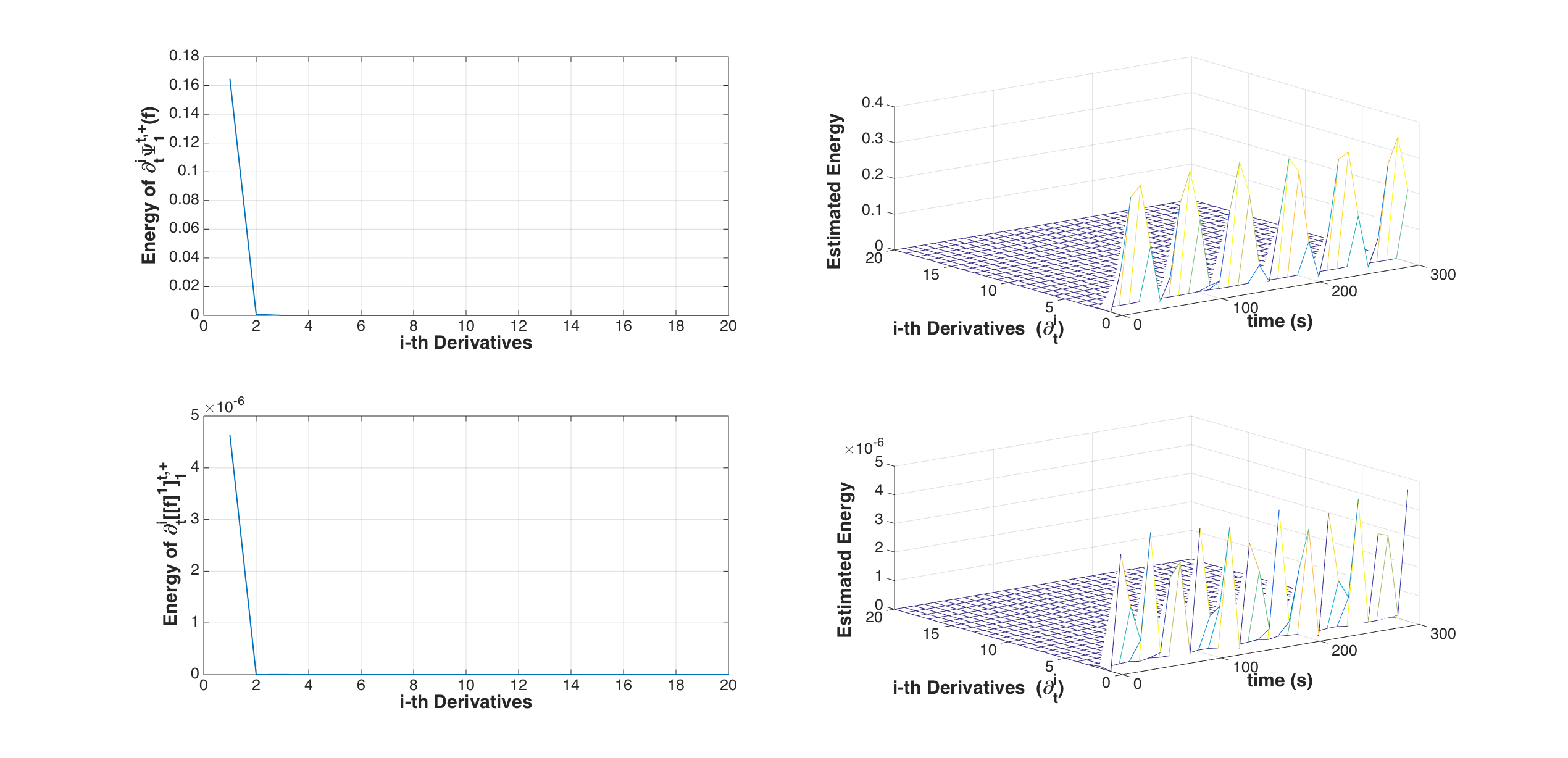}
\end{figure}
\begin{figure}\label{Fig1b}
\caption{ Results from the estimation of the energy of $\partial_r^i \Psi_1^{+,r} (f) (r,t_0)$   and $\partial_r^i [[f]^1]^{+,r}_1 (r,t_0))$.  The Figures correspond to Scenario  $2$.}
 \centering
 \includegraphics[width=1.3\textwidth]{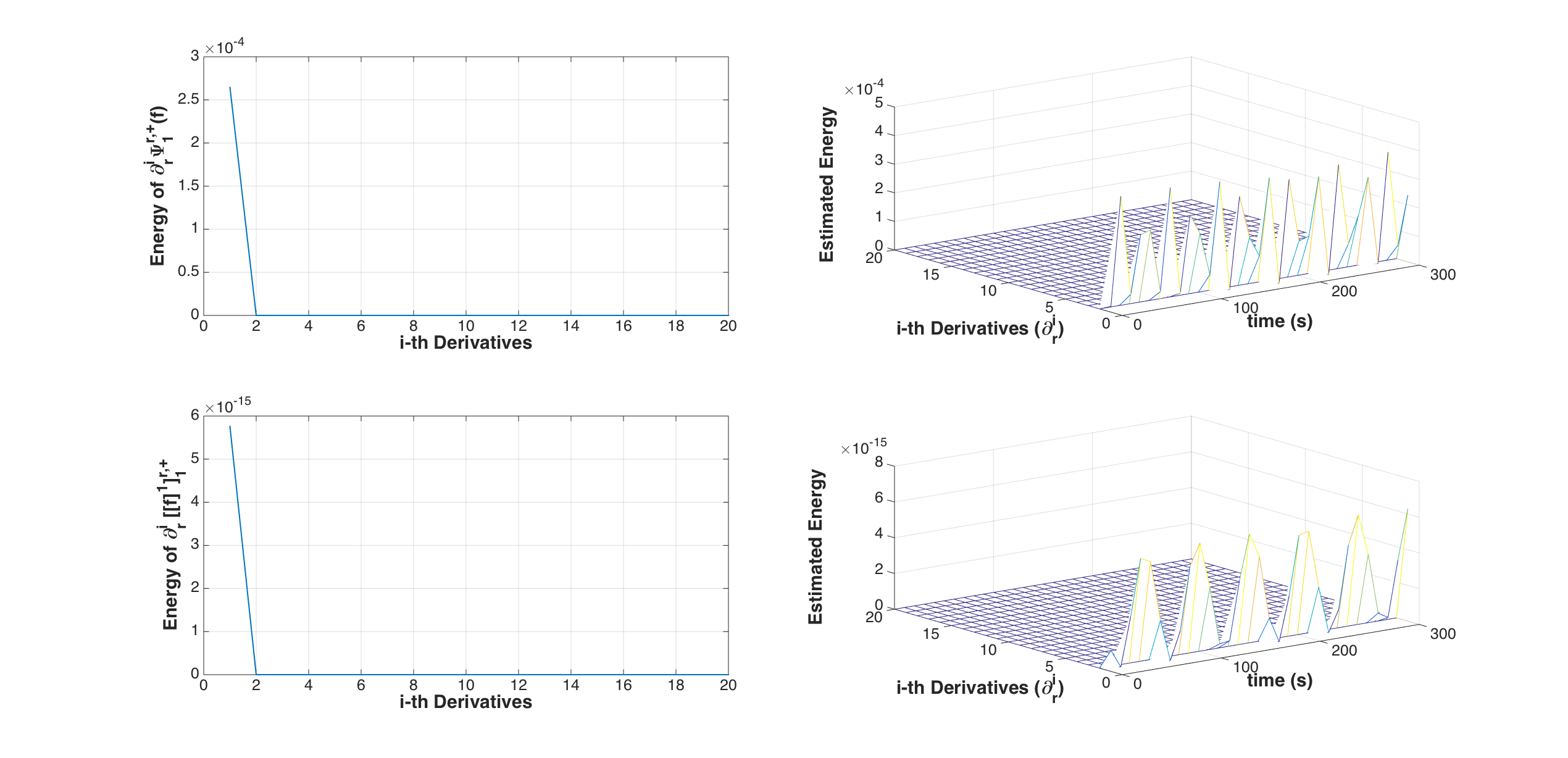}
\end{figure}
\newline The top figures in Figure $1$ and $2$ correspond to the numerical estimation of the energy of the operators $\partial_t^i \Psi_1^{+,t} (f) (r_0,t)$ (also called $\partial_t^i [[f]^0]^{+,t}_1 (r_0,t))$) and  f $\partial_r^i \Psi_1^{+,r} (f) (r,t_0)$ (also called $\partial_r^i [[f]^0]^{+,r}_1 (r,t_0))$). Thus, the results shows that for $i>2$, the values of $\mathcal{E}(\partial_t^i \Psi_1^{+,t} (f) (r_0,t))$ and $\mathcal{E}(\partial_r^i \Psi_1^{+,r}(f) (r,t_0))$ are smaller than $10^{-10}$. Thus, we can conclude that $i_1=i_2=3$. This simulation concludes the case study for the energy space $\mathbf{E}$ defined in Section $2.2$.
\\  Now, if we want to use the generalized energy operators (and $\bold{Theorem}$ $2$) with $p$ in $\mathbb{Z}^+$ and $p>0$, then one can look at solutions in the energy space $\mathbf{E}_p$ (e.g, Section $2.2$). It is the solutions of the kind $([[f]^p]^{+,t}_1)^n$ or $([[f]^p]^{+,r}_1)^n$. Thus, the definition of the subspaces  $\mathbf{N}_1^i$ and $\mathbf{N}_2^i$ can be extend with $p$ in $\mathbb{Z}^+$ such as
\begin{eqnarray}\label{energyspaceH002funB}
\mathbf{N}_{1,p}^i &=&\{ g \in \mathbf{S}^{-}(\mathbb{R}^2),\hspace{0.5em} \forall t \in [0, T ], \hspace{0.5em} r_0  \in [r_1, r_2] | g(r_0,t) = \partial_t^i ([[f]^p]^{+,t}_1)^n(r_0,t),  \nonumber \\
&& \hspace{0.5em} f  \in \mathbf{s}_p^{-}(\mathbb{R}^2),  \hspace{0.5em}n \in \mathbb{Z}^+ -\{0,1\} \} \nonumber \\
%
%
%
{\mathbf{N}}_{2,p}^i &=&\{  g \in \mathbf{S}^{-}(\mathbb{R}^2), \forall r   \in [r_1, r_2], \hspace{0.5em} t_0  \in [0, T ] | g(r,t_0) = \partial_r^i ([[f]^p]^{+,r}_1)^n(r,t_0), \nonumber \\
&& \hspace{0.5em} f  \in \mathbf{s}_p^{-}(\mathbb{R}^2), \hspace{0.5em} n \in \mathbb{Z}^+ -\{0,1\} \} \nonumber 
\end{eqnarray} 
With this definition, let us do a similar numerical estimation of the solutions and finding $i_1$ and $i_2$ for the case $p=1$. The bottom figures of Figure $1$ and $2$ are the results. We can see that there is a factor of $10^{-4}$ and $10^-{11}$  in Scenario $1$ and $2$ respectively between the estimated energy for the operators  corresponding to the case $p=0$ and $p=1$. Thus, one can conclude that $i_1$ and $i_2$ are equal to $0$. Recalling the decomposition  $\bold{Theorem}$ $1$ (and equation \eqref{energyspaceH22vxd}), the energy space $\mathbf{E}_0$ is reduced to $\{0\}$.  
\subsection{Phenomenon of Duplication of Waves in a Closed Cavity}\label{waveguideappli}
This section reflects on the idea of propagating electromagnetic waves in a closed cavity \cite{Annino}.  Here, the idea is not to revise the whole theory of electromagnetism, but rather discussing at a theoretical level the implication of our proposed model in this example and defining the $\it{duplication}$ of waves.
\\ Let us consider the form of solutions which propagates in a closed cavity. According to \cite{Petit}, one possible solution is the evanescent wave described in  equation \eqref{evanescent}. One can notice that $f$ and $\mathcal{E}(f)$ are analytic by definition for the variable $t$ and  $r$. With $\bold{Proposition}$ $1$,  we can  assume that $f$ is finite energy and in $\mathbf{s}^-(\mathbb{R}^2)$ with a wise choice on the parameters $A$, $k_1$, $k_2$ and $\omega$. One can estimate the difference of energy in time over $dt$ inside  the cavity at a specific location $r_0$ ($r_0$ in $[r_1, r_2]$) such as
\begin{eqnarray}\label{refeq100}
\mathcal{E}(f(r_0,T)) & =& \int_0^{T} (f(r_0,u))^2du < \infty \nonumber \\
\mathcal{E}(f(r_0,T+dt)) &=& \mathcal{E}(f(r_0,T)) + \sum_{k=0}^\infty \partial_t^k (f^2(r_0,T)) \frac{(dt)^k}{k!} <\infty \nonumber \\
\mathcal{E}(f(r_0,T+dt)) &=& \mathcal{E}(f(r_0,T)) + \sum_{k=0}^\infty \partial_t^k (f^2(r_0,T)) \frac{(dt)^k}{k!} <\infty \nonumber \\
\mathcal{E}(f(r_0,T+dt)) &=& \mathcal{E}(f(r_0,T)) + f^2(r_0,T)dt + \sum_{k=1}^\infty \partial_t^{k-1} \big (\Psi_1^{+,t}(f)(r_0,T) \nonumber \\
&& +\Psi_1^{-,t}(f)(r_0,T) \big) \frac{(dt)^{k+1}}{{k+1}!} <\infty \nonumber \\
\mathcal{E}(f(r_0,T+dt)) & \simeq & \mathcal{E}(f(r_0,T)) + f^2(r_0,T)dt 
\end{eqnarray}
Note that by definition $\Psi_1^{-,t}(f)(r_0,T) = 0$. Here the symbol '$\simeq$' means that 
%
\begin{equation}\label{approximate01}
\exists \hspace{0.5em} \epsilon \in \mathbb{R}^+, \hspace{0.5em} \epsilon <<1, \hspace{0.5em} \forall k \in \mathbb{Z}^+, \hspace{0.5em} k>0 | \hspace{0.5em}    |\partial_t^{k-1} \big (\Psi_1^{+,t}(f)(r_0,T)  \big ) | < \epsilon |f^2(r_0,T)|
\end{equation}
Now, let us do a hypothesis that $\mathcal{E}(f(r_0,T+dt))$ increases significantly over $dt$ modifying the approximation in \eqref{approximate01}
\begin{equation}\label{approximate01p}
\exists \hspace{0.5em} \epsilon \in \mathbb{R}^+, \hspace{0.5em} \epsilon <<1, \hspace{0.5em} \forall k \in \mathbb{Z}^+, \hspace{0.5em} k>1 | \hspace{0.5em}    |\partial_t^{k-1} \Psi_1^{+,t}(f)(r_0,T) | < \epsilon |\Psi_1^{+,t}(f)(r_0,T)|
\end{equation}
and then,
 \begin{eqnarray}
 \mathcal{E}(f(r_0,T+dt)) & \simeq & \mathcal{E}(f(r_0,T)) + f^2(r_0,T)dt + \Psi_1^{+,t}(f)(r_0,T) \frac{dt^2}{2} \nonumber \\
 \end{eqnarray}
Now, using the $\bold{Theorem}$ $1$ and the model based on the energy space in the previous sections, one can say that the subspace $\bold{N}_1^1$ defined in Section \ref{defsectionlPDE} is not reduced to $\{0\}$. Thus, the solutions from this subspace has to be taken into account. The $\it{duplication}$ of wave can be formulated as an approximation for taking into account additional solutions produced by the wave equation. 

\section{Conclusions}
The core of this work is to define the notion of $\it{multiplicity}$ of the solutions of a linear PDE using the model associated with energy spaces and the (generalized) energy operators. In this way, it contradicts the classical way of solving a nominated PDE with boundaries conditions, but it rather focuses on additional solutions from these energy spaces. The $\it{multiplicity}$ is defined through $\bold{Theorem}$ $2$ and the $\bold{Corollary}$ $1$. The work shows how the energy operators (and generalized energy operators) can determine which energy subspace is reduced to $\{0\}$. 
\\ The theory is then applied to the evanescent waves, a special type of solutions of the wave equation. The last part with the closed cavity show a possible real world applications. In this case, the $\it{duplication}$ of waves is when additional solutions should be taken into account due to the level of energy increasing in the cavity. However, this work remains at a theoretical level and more work with simulations are required to fully understand the concept of $\it{duplication}$.
\\ Furthermore, it was explained in Section \ref{nonlinearpdes} that the linearity of the PDEs does not play any role  in $\bold{Theorem}$ $2$ and the $\bold{Corollary}$ $1$. As a matter of fact, the model with the energy subspace is applied directly to the form of the solutions of the nominated PDE (i.e. evanescent waves solution of the wave equation) with the conditions that the solutions are finite energy and in the Schwartz space $\bold{S}^-(\mathbb{R}^2)$. Thus, there is no restriction theoretically speaking to use this model with non-linear PDEs. Our next interest is to apply our model to the type of solutions of the Korteweg de Vries equation called solitons which filled in the properties required to apply this model (e.g, finite energy function, solution in the Schwartz space $\bold{S}^-(\mathbb{R}^2)$ ).
\\ To conclude, the multiplicity of the solutions applied to the wave equation is only the first step before wondering if the receiving waves in a transmitter-receiver system, can be a sum of all those solutions. In other words, one can ask if the signal is generated by receiving not one type of solution/wave, but the additional solutions/waves coming from other energy subspaces.
\subsection*{Acknowledgments}
The author wants to dedicate his work to vale Emeritus Prof. Alan G.R. McIntosh (FAA) former member of the center for mathematics and its applications in the school of mathematical sciences at the Australian National University. He kindly introduced me to the Schwartz space theory when many years ago I started the development of the mathematical frame on energy operators.

 The authors would like also to thank the anonymous referee for his/her comments that helped us improve this article.

\appendix

\section{Discussion about $n$ in $\mathbb{Z}$}
Here is the discussion written in  \cite{JPMontillet2014} (Section 4) in the case of the generalized energy operators about extending  $\bold{Theorem}$ $1$ and $\bold{Lemma}$ $1$ with $n$ in $\mathbb{Z}$. The extension of $\bold{Theorem}$ $0$ and $\bold{Lemma}$ $0$ can be found in \cite{JPMontillet2013} (Section 3, p. 74). Let us recall:
\vspace{0.5em}
\\$\bold{Discussion}$ $n<-1$: In this case, one can define:
\begin{equation}\label{setbbb}
\forall f \in \mathbf{S}_p^{-}(\mathbb{R}), \hspace{0.5em} \forall t \in \mathbb{R}, \hspace{0.5em} p \in \mathbb{Z}^+, \hspace{0.5em} ([[f(t)]^p]_1^+\big )^n\neq 0, \hspace{0.5em} \forall n \in \mathbb{Z}^+, n>1, \frac{1}{\big ([[f(t)]^p]_1^+\big )^n}
\end{equation}
This set of functions can also be described as: $f$ in $\mathbf{S}_p^{-}(\mathbb{R})$ and $f$ not in  $Ker\big ([[f(t)]^p]_1^+\big )$ for $p$ in $\mathbb{Z}^+$. Note that one could also chose to have $f$ in $\mathbf{s}_p^{-}(\mathbb{R})$. However, this is more restrictive than the set defined in \eqref{setbbb}. Using an intermediary function, $h$ such as $h = \frac{1}{[[f(t)]^p]_1^+}$, the problem of decomposing $\partial_t^k \big ([[f(t)]^p]_1^+\big )^{-n}$ ($k$ in $\mathbb{Z}^+-\{0\}$) is equivalent to resolving $\partial_t^k h^{n}$, which has been demonstrated in the $\bold{Lemma}$ $1$ and $\bold{Theorem}$ $1$.
\vspace{1.0em}
\newline $\bold{Discussion}$ $n=1$ or $n=-1$: As already underlined in \cite{JPMontillet2013}, one can use a general formula for $f$ in the set defined in equation \eqref{setbbb}:
\begin{eqnarray}\label{discussion2a}
\partial_t^k \big ([[f(t)]^p]_1^+\big ) &=& \partial_t^k \bigg (\frac{\big ([[f(t)]^p]_1^+\big )^3}{\big ([[f(t)]^p]_1^+\big )^2}\bigg) \nonumber \\
k=1, \qquad \partial_t \big ([[f(t)]^p]_1^+\big ) &=& \big ([[f(t)]^p]_1^+\big )^{-2} \partial_t \big ([[f(t)]^p]_1^+\big )^3 + \big ([[f(t)]^p]_1^+\big )^3 \partial_t \big ([[f(t)]^p]_1^+\big )^{-2} \nonumber \\
k=2, \qquad \partial_t^2 \big ([[f(t)]^p]_1^+\big ) &=& 2 \partial_t \big ([[f(t)]^p]_1^+\big )^{-2} \partial_t \big ([[f(t)]^p]_1^+\big )^3 \nonumber \\
&& + \big ([[f(t)]^p]_1^+\big )^3 \partial_t^2 \big ([[f(t)]^p]_1^+\big )^{-2} + \big ([[f(t)]^p]_1^+\big )^{-2} \partial_t^2 \big ([[f(t)]^p]_1^+\big )^3 \nonumber \\
\end{eqnarray} 
The example for $k =\{1,2\}$ in Equation \eqref{discussion2a} shows that $\partial_t^k \big ([[f(t)]^p]_1^+\big )$  can be decomposed into a product of successive derivatives of $\big ([[f(t)]^p]_1^+\big )^{3}$ and $ \big ([[f(t)]^p]_1^+\big )^{-2}$. Those derivatives can be decomposed into a sum of generalized energy operators based on the $\bold{Lemma}$ $1$ and $\bold{Theorem}$ $1$ plus the previous discussion (for the case $n<-1$ ).
\\ Now for the case $n=-1$, it is easy to see that :
\begin{eqnarray}\label{discussion2amoins1}
\partial_t^k \big ([[f(t)]^p]_1^+\big )^{-1} &=& \partial_t^k \bigg (\frac{\big ([[f(t)]^p]_1^+\big )^2}{\big ([[f(t)]^p]_1^+\big )^3}\bigg) \nonumber \\
\end{eqnarray}
With the discussion for the case $n=1$, we can conclude that $\partial_t^k \big ([[f(t)]^p]_1^+\big )^{-1}$  can be decomposed into a product of successive derivatives of $\big ([[f(t)]^p]_1^+\big )^{2}$ and $ \big ([[f(t)]^p]_1^+\big )^{-3}$.
\end{document}